\documentclass[12pt]{article}

\usepackage{amssymb}

\begin{document}

\begin{titlepage}

\begin{flushright}
\end{flushright}
\vskip 2.5cm

\begin{center}
{\Large \bf Testing Electron Boost Invariance with\\
$2S$-$1S$ Hydrogen Spectroscopy}
\end{center}

\vspace{1ex}

\begin{center}
{\large Brett Altschul\footnote{{\tt baltschu@physics.sc.edu}}}

\vspace{5mm}
{\sl Department of Physics and Astronomy} \\
{\sl University of South Carolina} \\
{\sl Columbia, SC 29208} \\
\end{center}

\vspace{2.5ex}

\medskip

\centerline {\bf Abstract}

\bigskip

There are few good direct laboratory tests of boost invariance for electrons, because
the experiments required often involve repeated
precision measurements performed at different times of year. However, existing
measurements and remeasurements of the $2S$-$1S$ two-photon transition frequency in 
$^{1}$H---which were done to search for a time variation in the fine structure
constant---also constitute a measurement of the boost symmetry violation parameter $0.83c_{(TX)}+0.51c_{(TY)}+0.22c_{(TZ)}=(4\pm8)\times 10^{-11}$. This is an eight
order of magnitude improvement over preexisting laboratory bounds, and with only one
additional measurements, this system could yield a second comparable constraint.

\bigskip

\end{titlepage}

\newpage

There is currently a great deal of interest in the possibility that Lorentz and
$CPT$ symmetries may not be exact in nature~\cite{ref-review1}. A diverse spectrum of
precision experiments have placed bounds on many different forms of Lorentz violation.
These various forms are described by the parameters of the standard model extension
(SME), an effective quantum field theory~\cite{ref-kost2}.
Bounds on SME parameters are summarized in~\cite{ref-tables}.

Atomic spectroscopy is a powerful tool in precision physics. Many of the most
precise tests of Lorentz symmetry have used atomic
clocks~\cite{ref-berglund,ref-kost6,ref-humphrey,ref-cane,ref-wolf,ref-kornack}. These
experiments mostly involve measurements of nuclear spin transitions, and less
attention has been paid to
optical transitions, which are typically sensitive to different sets SME
coefficients. Optical transition frequencies are generally more sensitive to Lorentz
violation in the electron sector and to forms of Lorentz violation that are
independent of particle spin.

We shall show how existing optical spectroscopy measurements can be used to place new
bounds on Lorentz violation. The measurements involved were of the $2S$-$1S$
two-photon transition frequency in hydrogen~\cite{ref-niering,ref-fischer}. This has
historically been one of the best known optical transition frequencies, and it
provides useful information about many different physical phenomena.
Lorentz violation in $^{1}$H was previously discussed in~\cite{ref-bluhm3}. However,
since
the focus there was primarily on frequency shifts that were not suppressed by any
power of the fine structure constant $\alpha$, the class of experiments described here
was not discussed.

The electron Lagrange density relevant for the $^{1}$H experiments is
\begin{equation}
\label{eq-L}
{\cal L}=\bar{\psi}[(\gamma^{\mu}+c^{\nu\mu}\gamma_{\nu})
i\partial_{\mu}-m]\psi,
\end{equation}
where $c^{\nu\mu}$ is a traceless
background tensor describing the Lorentz violation. The
electron sector of the SME contains other tensors, which parameterize different types
of Lorentz and $CPT$ violation. However, most of these depend on spin, and the ones
that could affect the energy difference between two atomic $S$ states are already
strongly constrained by torsion pendulum experiments with spin-polarized
samples~\cite{ref-heckel2}.

The $c_{00}$ term in ${\cal L}$ enters as a rescaling of the time derivative term
in the action. Under quantization, $i\partial_{0}$ becomes the electron energy $E$,
and $c_{00}$ simply rescales all electronic energy eigenvalues.
Including the effects of the all the $c^{\nu\mu}$ coefficients,
the nonrelativistic Hamiltonian for the electron is~\cite{ref-kost10}
\begin{equation}
H=\frac{p_{j}p_{k}}{2m}\left[\delta_{jk}-c_{00}\delta_{jk}-c_{(jk)}\right]
-c_{(0j)}p_{j}+V.
\end{equation}
The coefficients $c_{(\nu\mu)}$ are the symmetrized combinations
$c_{\nu\mu}+c_{\mu\nu}$.

Bounds on SME coefficients are generally given in sun-centered
celestial equatorial coordinates~\cite{ref-bluhm4}. The sun-centered reference
frame is approximately inertial on all relevant time scales, and it provides a
convenient way to parameterize measurements of boost symmetry violation that make
use of the Earth's orbital motion (such as the measurements discussed here).
The Cartesian coordinates used in this frame are $(X,Y,Z,T)$.
The spatial origin lies at the center of the sun.
The $Z$-axis points along the direction of the Earth's rotation, and the $X$-axis
points toward the vernal equinox point on the celestial sphere. (This means Earth lies on the negative $X$-axis at the time of this equinox.) The
$Y$-direction is chosen according to the right hand rule, and the origin of time
($T=0$) is taken to be the vernal equinox in the year 2000.

There are no strong laboratory bounds on the $c_{(TJ)}$. Constraints based on
Doppler effect measurements are only at the $10^{-2}$ level~\cite{ref-lane1};
and even these weak bounds are only order of magnitude inferences based on earlier
experimental data. The experiment used spectroscopy of
$^{7}$Li$^{+}$ ions moving at a speed $v\approx0.064$ to test the accuracy of the
relativistic Doppler shift formula. The experiment confirmed the conventional
prediction with $\sim 2\times10^{-9}$ precision~\cite{ref-saathoff}, providing a weak
sensitivity to the electron $c_{(TJ)}$ coefficients.

There are better astrophysical bounds~\cite{ref-altschul7}, but they have a number of
fairly undesirable features. These bounds are derived from observations of
extremely high energy astrophysical phenomena. In order to translate the
observational data into constraints on the $c_{(TJ)}$ and other
coefficients, we must understand how the energetic phenomena really work.
For example, the TeV $\gamma$-rays from many sources are probably produced
by the upscattering of low-energy photons by extremely energetic electrons
(inverse Compton scattering), and the photon spectrum can tell us a great
deal about the behavior of the highly boosted electrons. However, there is
an alternative hypothesis for the origins of the TeV photons---that they
are produced in $\pi^{0}$ decay; if that is the case, then observations of
the photons may really tell us very little about electron Lorentz
violation.

Yet there is one type of astrophysical bound that is immune to this
problem.  However TeV $\gamma$-rays are produced, they could potentially decay
($\gamma\rightarrow e^{+}+e^{-}$) if there is Lorentz violation in the
electron sector. The observed absence of this process allows us to place
some stringent constraints on the $c_{(TJ)}$ and $c_{TT}$
coefficients~\cite{ref-altschul14}.
However, these constraints are all one-sided, and it is impossible to
disentangle the $c_{(TJ)}$ bounds from those on $c_{TT}$.
Consequently, these measurements cannot
exclude any specific values of $c_{(TJ)}$. This makes it extremely important
to have separate, two-sided bounds on the $c_{(TJ)}$.

$c_{00}$ acts to shift all electron energies, and its effects may be
seen using any electron transition. Moreover, if one measures an
observable that is even under $C$, $T$, and any reflection (the parity
operator $P$ may be decomposed into three separate reflections
$P=R_{1}R_{2}R_{3}$), $c_{00}$ and $c_{jk}$ with $j=k$ are the only electron
coefficient that can contribute at first order.
All other forms of electron-sector Lorentz violation are odd under at
least one of $C$, $T$, or $R_{j}$. The anisotropic $c_{JK}$ coefficients can be
constrained by looking at Michelson-Morley experiments with material-filled
cavities~\cite{ref-muller2}, and so we shall neglect them here.

The $c_{00}$ to which the $1S$-$2S$ two-photon transition frequency is
sensitive is the coefficient in the laboratory frame. In terms of the
sun-centered coefficients, $c_{00}=c_{TT}+v_{J}c_{(TJ)}$,
to first order in the velocity $\vec{v}$ of the laboratory. The $c_{TT}$
contribution is uniform across all reference frames and therefore difficult to
measure. However, it has been bounded at the $10^{-15}$ level
using data from accelerator
experiments~\cite{ref-hohensee1,ref-altschul20}, which are more
sensitive to violations of boost invariance. For these reasons, $c_{TT}$ should
be unimportant, and we shall henceforth neglect it.

Since the contribution $c_{(TJ)}$ makes to $c_{00}$ depends on the rest frame of the
experiment, at different positions along the Earth's orbit the $c_{(TJ)}$ will affect
atomic energy levels differently. Year-round measurements would make it
possible to sample both $c_{(TJ)}$ corresponding to directions $\hat{J}$ lying
in the ecliptic plane. Schematically, this means measuring a $P$-even observable in
the laboratory frame, then comparing the measured values of this observable in
laboratories moving with different velocities. The differences in measured values
are not invariant under $P$ and so are sensitive to the $c_{(TJ)}$,
which are $P$-odd in the sun-centered frame.

The $c_{(TJ)}$ parameters describe violations of boost invariance. Testing
this invariance requires a comparison of the physics in two different
frames, and the effects of $c_{(TJ)}$ are suppressed by one power of the
velocity difference between the frames being compared. In contrast, the
isotropic boost invariance violation coefficient $c_{TT}$ can only
generate differences in physical observables between the two frames that
are suppressed by two powers of the relative speed. It is probably because
$c_{TT}$ thus seems substantially more difficult to constrain on its own that the
current methodology has not been much discussed. As well as mixing with
$c_{00}$, the $c_{(TJ)}$ mix with the laboratory $c_{jk}$ coefficients, which are
relatively easy to bound using a rotating apparatus (or just the rotating Earth), and
many attempts to
constrain boost invariance violation have looked for differences in the
anisotropy observed in different laboratory frames.
If the pure anisotropy coefficients $c_{JK}$ in the sun-centered frame are zero, then
$c_{(jk)}=v_{j}c_{(Tk)}+v_{k}c_{(Tj)}$ in the laboratory frame.
However, measurement of the anisotropy associated with this $c_{jk}$ is not
necessarily the best way to measure boost invariance
violation. The mixings of the $c_{(TJ)}$ with both the $c_{00}$ and
$c_{jk}$ generate observable effects at first order in $\vec{v}$.

The $^{1}$H atom is an extremely clean system, containing only an
electron and a proton. This simplifies theoretical calculations of energy
eigenvalues, and it makes it easier to extract bounds on electron-sector
coefficients; there are no many-body effects to complicate things.
The presence of $c^{\nu\mu}$ shifts the energy of a $nS$ state by
\begin{equation}
\Delta E=-\frac{m\alpha^{2}}{2n^{2}}\left(c_{00}+\frac{2}{3}c_{jj}\right);
\end{equation}
as expected,
only those laboratory-frame coefficients that are even under all discrete symmetries
contribute.
Moreover, because of the extremely long lifetime ($\approx 0.122$ s) of the $^{1}$H
$2S$ state, the $2S$-$1S$
two-photon transition frequency can be measured quite precisely. 
These facts make this system an extremely attractive place to search for evidence of
Lorentz violation.

Precisely the right kind of data for constraining the $c_{(TJ)}$ using
$^{1}$H spectrum measurements is already available. Recognizing the
usefulness of the $2S$-$1S$ $^{1}$H transition as a probe of
exotic physics, the laser spectroscopy group at the Max Planck Institute
made multiple measurements of its frequency. They used these measurements
to constrain the possible time variation of the fine structure constant
$\alpha$. However, their measurements, separated by approximately
44 months, are also almost ideally suited to constrain the electron
$c_{(TJ)}$, because the Earth's orbital velocities at the times of the two
measurements were distinctly different.

The frequency difference between measurements made at two different points along
the Earth's orbit is
\begin{equation}
\frac{\delta\nu}{\nu}=\frac{5}{3}\delta v_{J}c_{(TJ)},
\end{equation}
where $\delta\vec{v}=\vec{v}_{2}-\vec{v}_{1}$ is the velocity difference between the
two locations.
Earth's velocity in its orbit is (neglecting small effects such as the orbit's
ellipticity)
\begin{equation}
\vec{v}=v_{\oplus}\left(\hat{X}\sin\Omega_{\oplus}T-\hat{Y}\cos\eta
\cos\Omega_{\oplus}T-\hat{Z}\sin\eta\cos\Omega_{\oplus}T\right),
\end{equation}
where $v_{\oplus}\approx10^{-4}$ and $\Omega_{\oplus}$ are the speed and angular
frequency of the orbital revolution, and $\eta\approx 23.4^{\circ}$ is the
inclination between the equatorial and ecliptic planes~\cite{ref-kost16}.
The additional velocity due to planetary rotation is substantially smaller.

The actual measurements were made over periods of weeks, during which
time the velocity changed somewhat. In light of this fact, $\vec{v}_{1}$ and
$\vec{v}_{2}$ should be replaced by average velocities over the periods during which
the measurements were taken. If measurements are taken uniformly in time over a
period $\Delta T$, the average orbital velocity over this period is
$\frac{\sin(\Omega_{\oplus}\Delta T/2)}{\Omega_{\oplus}\Delta T/2}$ times the
velocity at the midpoint of the observation period. For a one-month observation
window, the averaging reduces the velocity by less than 5\%; for two months,
the reduction is less than 18\%. Averaging over June and July 1999 for the first
measurement and February 2003 for the second, the effective velocity difference
between the two observation periods is
\begin{equation}
\label{eq-deltav}
\delta\vec{v}=-v_{\oplus}(1.35\hat{X}+0.83\hat{Y}+0.36\hat{Z}).
\end{equation}
Measurements taken over short periods of time at opposite points on the orbit would
maximize $|\delta\vec{v}|$, but the estimated experimental value (\ref{eq-deltav})
differs from the ideal one by only about 22\%.

The observed difference $\delta\nu$ between the frequencies measured in the
two experiments was $(-29\pm57)$ Hz.
Compared with the $2S$-$1S$ frequency of $2.4660611024748\times 10^{15}$ Hz, this
represents a fractional difference of
$\frac{\delta\nu}{\nu}=(-1.2\pm2.4)\times10^{-14}$.
The resulting measurement of the electron $c_{(TJ)}$ coefficients is
\begin{equation}
0.83c_{(TX)}+0.51c_{(TY)}+0.22c_{(TZ)}=(4\pm8)\times 10^{-11}.
\end{equation}

Another precision measurement of the $2S$-$1S$ transition frequency $\nu$ would
constrain an independent linear combination of the $c_{(TJ)}$ parameters. For
optimal results, such a measurement should be performed in October or November, so
the three measurements are roughly evenly spaced around the
Earth's orbit. The linearly independent combination that would be bounded is
$0.56c_{(TX)}-0.76c_{(TY)}-0.33c_{(TZ)}$. However, the
third independent combination of SME coefficients---which is the component
of $c_{(TJ)}$ associated with boosts in the direction normal to the orbital
plane---cannot be constrained this way, because the
Earth's velocity in this direction
does not change. Laboratory measurements of this kind of boost invariance violation
would need to rely on the Earth's rotational velocity, which is about two orders of
magnitude smaller than $\vec{v}_{\oplus}$ but which would combine with the Lorentz
violation to produce sidereal variations in $\nu$.

The $2S$-$1S$ spectroscopy measurements used a $^{133}$Cs atomic clock as a
frequency standard. In principle, Lorentz violation in this nucleus could also
affect the experimental results. However, the relevant nuclear SME parameters
have already been strongly constrained~\cite{ref-wolf} by a comparison of
$^{133}$Cs atomic clocks with other frequency standards.
Ultimately, only differences between these frequency standards are observable, and
the $2S$-$1S$ spectroscopy provides a new constraint on the difference between
the electron sector and others.

Other experiments already performed could provide similar but less precise bounds
on the electron boost invariance violation coefficients. There are earlier, less
precise measurements of the $2S$-$1S$ frequency difference in
$^{1}$H~\cite{ref-udem}. There are also other recent measurements that
were spread out in time to search for changes in $\alpha$. Optical transitions in $^{199}$Hg$^{+}$~\cite{ref-bize1}, $^{171}$Yb$^{+}$~\cite{ref-peik},
and $^{87}$Sr~\cite{ref-blatt}
were studied over periods of years, with accuracies comparable to that of the $^{1}$H
experiment. These kinds of
experiments could be used to place similar bounds, although the analyses are not
so simple as with $^{1}$H, since the transitions are no longer between pairs of
isotropic states. The dependence on the laboratory $c_{jk}$ coefficients would
depend on the orientation of the apparatus, and the energies would be subject to
sidereal variations. The atomic structures are also more
complicated. Moreover, the $^{199}$Hg$^{+}$ and $^{171}$Yb$^{+}$ bounds were based on
measurements made
mostly around one time of year, which makes them less useful
for constraining annual variations. The Sr experiment looked explicitly for annual
variations, but the analysis assumed that frequency changes would be tied to
variations in the Earth-sun distance. Taken together, the data from all these
experiments suggest that both components of $c_{(TJ)}$ corresponding to directions
$\hat{J}$ in the orbital plane should be bounded at the $\lesssim10^{-9}$ level,
but further analysis would be required to make this bound firm.

The measurement and remeasurement of the $2S$-$1S$ two-photon transition
frequency $\nu$ was conceived as a way to measure the time variation of $\alpha$.
But many of the same properties of these measurements---their
high precision, relatively short duration, and substantial separation in time---that made them ideal for constraining $\dot{\alpha}$ also make them sensitive to the
Lorentz violation coefficients $c_{(TJ)}$. The experimental data constrain a
specific combination
of these coefficients at the $10^{-11}$ level; this represents an eight order of
magnitude improvement over previous laboratory constraints. Moreover, these
bounds could easily be extended to another combination of $c_{(TJ)}$ coefficients
simply by repeating the experimental measurement a third time.

\end{document}